\documentclass[aps,prd,reprint,preprintnumbers,showpacs,floatfix,nofootinbib,superscript address,longbibliography]{revtex4-1}
\usepackage[utf8]{inputenc}
\usepackage{amssymb}
\usepackage{hhline}
\usepackage{amsmath}
\usepackage{mathtools}
\usepackage[dvipsnames]{xcolor}
\usepackage{multirow,tabularx}
\usepackage{graphicx}
\usepackage{natbib}

\usepackage{xspace}
\usepackage{xstring}
\usepackage{titlesec}
\usepackage{parskip}
\allowdisplaybreaks
\usepackage{braket}

\newrobustcmd{\pea}[1]{%
	\emph{#1}\textbf{\ \ \ ---}
}
\titleformat{\paragraph}[runin]{\normalfont\normalsize\bfseries}{\emph\theparagraph}{1em}{\pea}

\newcommand*{\ie}{i.e.\@\xspace}
\newcommand*{\eg}{e.g.\@\xspace}
\newcommand*{\cf}{c.f.\@\xspace}

\newcommand*{\figs}{Figs.\@\xspace}
\newcommand*{\eq}{eq.\@\xspace}

\newcommand*{\ex}{\mathrm{e}}

\usepackage{hyperref}
\hypersetup{%
     colorlinks = true,%
     linkcolor = Blue,%
     citecolor = Blue,%
     filecolor = Blue,%
     urlcolor = Blue%
     }%

\begin{document}

\title{Black Hole Memory Burden and its Signatures in Gravitational Waves from Mergers}

\author{Gia Dvali}
\email{gdvali@mpp.mpg.de}
\affiliation{Arnold Sommerfeld Center, Ludwig-Maximilians-Universit{\"a}t, Theresienstr.~37, 80333 M{\"u}nchen, Germany}
\affiliation{Max-Planck-Institut f{\"u}r Physik, Boltzmannstr.~8, 85748 Garching, Germany}

\author{Michael Zantedeschi}
\email{michael.zantedeschi@pi.infn.it}
\affiliation{INFN, Sezione di Pisa,
Largo Bruno Pontecorvo 3, I-56127 Pisa, Italy}

\author{Sebastian Zell}
\email{sebastian.zell@lmu.de}
\affiliation{Arnold Sommerfeld Center, Ludwig-Maximilians-Universit{\"a}t, Theresienstr.~37, 80333 M{\"u}nchen, Germany}
\affiliation{Max-Planck-Institut f{\"u}r Physik, Boltzmannstr.~8, 85748 Garching, Germany}

\begin{abstract}
Swift \emph{memory burden} (MB) implies that the information stored in a black hole (BH) can modify its classical dynamics when the BH is perturbed. This influences the gravitational waves (GWs) emitted during BH mergers. 
   In this paper, we investigate how the BH memory load is determined by the features of the collapsing source.  We show that the memory load can vastly exceed the information content of its progenitor. An extreme example is a BH formed in a two-particle collision, which exhibits maximal MB.
   We then derive bounds for BHs formed through stellar collapse and examine the impact of swift MB on BH quasinormal modes, quantifying the MB-induced frequency shift of GWs. 
     These findings imply that GW observations probe the fundamental mechanisms of BH information storage as well as their formation history.
\end{abstract}

\maketitle

\tableofcontents

\section{Introduction}

One of the most remarkable properties of black holes (BHs) is their ability to store an enormous amount of information, as quantified by their large Bekenstein–Hawking entropy~\cite{Bekenstein:1973ur,Hawking:1975vcx}. 
Nevertheless, in standard treatment the role of 
a BH's information in its macroscopic dynamics has not been taken into account. 

This attitude has been challenged by the discovery of  the \emph{memory burden (MB)} effect~\cite{Dvali:2018xpy,Dvali:2018ytn,Dvali:2020wft,
Dvali:2021tez, Alexandre:2024nuo, Dvali:2024hsb}.
 It makes a clear distinction between the entropy which measures the capacity of information storage and the actual load of information (memory) carried by a particular BH. 
The MB phenomenon also quantifies the influence of BH information in terms of an invariant notion of the memory load accounted by the the fraction of the used memory space.
 The MB effect shows that the memory load strongly influences the  dynamics of a BH and in particular tends to stabilize it against Hawking decay.  

However, the backreaction from the MB effect is not limited to evaporation. Rather, it can be caused by an arbitrary perturbation that takes the BH away from its ground state -- this is the \emph{swift MB (SMB)} effect~\cite{Dvali:2025sog}. 
The key point of SMB is that classically-identical BHs, not only age differently under a quantum decay, but also respond very differently to classical perturbations. 
In this sense,  SMB makes the role of BH  information load visible also in classical evolution.
As a direct consequence, SMB is expected to affect the gravitational waves (GWs) emitted in BH mergers, in particular, via a shift in the frequency of the emitted radiation~\cite{Dvali:2025sog}. 

Since the groundbreaking first detection of GWs~\cite{LIGOScientific:2016aoc}, the catalog of confirmed GW events has expanded rapidly~\cite{LIGOScientific:2018mvr,LIGOScientific:2020ibl,KAGRA:2021vkt,LIGOScientific:2025slb}, culminating this year in the loudest event observed to date, GW250114~\cite{LIGOScientific:2025rid,LIGOScientific:2025obp}. These observations have opened an entirely new window onto strong-field gravity and the fundamental physics of BHs~\cite{LIGOScientific:2016lio,LIGOScientific:2018dkp,LIGOScientific:2019fpa,LIGOScientific:2020tif,LIGOScientific:2021sio,LIGOScientific:2025rid,LIGOScientific:2025obp}. The last phase of a BH merger -- the ringdown -- is characterized by exponentially damped oscillations of the remnant BH, known as \emph{quasinormal modes (QNMs)}~\cite{Regge:1957td,Zerilli:1970wzz,Vishveshwara:1970zz,Press:1971wr}. 

Observations of these QNMs can test BH properties through black hole spectroscopy~\cite{Dreyer:2003bv,Gossan:2011ha} (for a review, see~\cite{Berti:2025hly}). A recent study~\cite{Yuan:2025hls} has applied this framework to probe SMB. In this paper, we shall clarify and extend the predictions of SMB~\cite{Dvali:2025sog} for BH QNMs.

The MB effect~\cite{Dvali:2018xpy,Dvali:2018ytn,Dvali:2020wft} arises generically in systems with efficient information storage capacity since all such systems produce large number of gapless memory modes that store information~\cite{Dvali:2017ktv,Dvali:2017nis,Dvali:2018vvx,Dvali:2018xoc,Dvali:2018tqi}.  In particular, the effect has been demonstrated for many examples of solitons~\cite{Dvali:2021bsy,Dvali:2021tez,Dvali:2023xfz,Dvali:2024hsb,Dvali:2025ktz,Contri:2025eod,Dvali:2025sog,Dvali:2025wto}.  The phenomenon of MB is 
especially prominently represented in objects that saturate the field theoretic bound on microstate degeneracy~\cite{Dvali:2020wqi} since such systems attain the maximal efficiency of memory storage.

The origin of MB effect can be traced to the fact that 
the stored information load creates an energetic barrier that inhibits departure from the initial state. MB is a universal effect and is particularly relevant for BHs due to their enormous entropy $S$~\cite{Bekenstein:1973ur,Hawking:1975vcx}. For instance, a BH of roughly $30$ solar masses -- as in the event GW250114~\cite{LIGOScientific:2025rid,LIGOScientific:2025obp} -- has $S \sim 10^{80}$, corresponding to approximately $\exp(10^{80})$ distinct possible microstates.
Although these microstates are degenerate in the BH's ground state, they carry vastly different memory loads. Therefore, the BH's evolution is sensitive to the  
initial microstate and is very different depending on its memory load.

One of the effects of MB is that it significantly slows down the BH evaporation~\cite{Dvali:2020wft}. This has important implications for primordial black holes (PBHs)~\cite{Zeldovich:1967lct,Hawking:1971ei,Carr:1974nx,Chapline:1975ojl,Carr:1975qj} formed in the early Universe. If semi-classical evaporation remained valid throughout the lifetime of a BH, small PBHs with masses below $10^{15}\,\text{g}$ would have fully evaporated by today (see \eg~\cite{Carr:2020xqk,Green:2020jor} for reviews). However, if MB slows evaporation, PBHs lighter than $10^{15}\,\text{g}$ can survive until the present epoch and thus constitute viable dark matter candidates~\cite{Dvali:2020wft, Alexandre:2024nuo, Thoss:2024hsr, Dvali:2024hsb}. Consequently, small PBHs make it possible to test MB through astrophysical observations~\cite{Dvali:2021byy,Franciolini:2023osw, Alexandre:2024nuo,Thoss:2024hsr,Dvali:2024hsb,Balaji:2024hpu,Haque:2024eyh,Barman:2024iht,Bhaumik:2024qzd,Barman:2024ufm,Kohri:2024qpd,Borah:2024bcr,Chianese:2024rsn,Zantedeschi:2024ram,Barker:2024mpz,Jiang:2024aju,Loc:2024qbz,Basumatary:2024uwo,Federico:2024fyt,Athron:2024fcj,Barman:2024kfj,Bandyopadhyay:2025ast,Calabrese:2025sfh,Boccia:2025hpm,Liu:2025vpz,Dvali:2025ktz,Montefalcone:2025akm,Chianese:2025wrk,Chaudhuri:2025rcs,Tan:2025vxp,Dondarini:2025ktz,Ettengruber:2025kzw,Chaudhuri:2025asm,Kitabayashi:2025iaq,Maity:2025ffa,Thoss:2025yht,Gross:2025hia,Merchand:2025bzt,Tseng:2025fjf,Levy:2025lyj,Sarmah:2025wrg,Ambrosone:2026djo,Ettengruber:2026jrc,Ewasiuk:2026ncb,Thoss:2026slt,Chaudhuri:2026ljw,Leontaris:2026kvu,Ahmed:2026pjd,Du:2026vxw,Zhang:2026dkq}, with important signatures in GWs~\cite{Franciolini:2023osw,Balaji:2024hpu,Barman:2024iht,Bhaumik:2024qzd,Barman:2024ufm,Kohri:2024qpd,Barker:2024mpz,Jiang:2024aju,Loc:2024qbz,Basumatary:2024uwo,Athron:2024fcj,Dondarini:2025ktz,Gross:2025hia,Merchand:2025bzt,Ahmed:2026pjd} and high-energy astroparticles~\cite{Thoss:2024hsr,Chianese:2024rsn,Zantedeschi:2024ram,Liu:2025vpz,Boccia:2025hpm,Dvali:2025ktz,Chianese:2025wrk,Tan:2025vxp,Dondarini:2025ktz,Tseng:2025fjf,Ambrosone:2026djo,Chaudhuri:2026ljw}.

While BH evaporation is a quantum process, the SMB influences the classical perturbations of a BH affecting, among others, the GWs emitted in BH mergers~\cite{Dvali:2025sog}.
 In particular,  for weak-perturbations, $|\delta g| \ll 1$, the coherent mode experiences a relative shift of the resonant frequency of GWs emitted during the inspiral phase, \begin{equation} \label{DeltaF1}
     |\Delta f/f| \simeq \mu^{-1} |\delta g|^{2(p-1)}\,,
 \end{equation}
 where $\mu$ is the MB parameter that measures the relative weight of the information load and $p$ is the critical exponent. 
 
  The  goal of the present paper is to elucidate the observational prospects of probing the  BH information storage mechanism via the above formula.  Of course, the
  key question is to understand how accurately $\mu$ can be predicted for a BH formed by the collapse of a given astrophysical source. Some preliminary estimates were 
  already performed in~\cite{Dvali:2025sog}, on which  we shall improve. 
  We show that 
$\mu$ is determined both by the diversity of the features of the collapsing source as well as by the microscopic mechanism that encodes these features in BH memory modes. 
Thus, the connection between the features 
of the source and $\mu$ is rather profound and other factors must be taken into account.  These factors allow 
 almost featureless sources to carry maximal MB and vice versa. 
For example, based on symmetry and unitarity arguments, we  conclude that a macroscopic BH formed by a featureless microscopic source  carries a maximal MB. 
This is due to a particular manifestation of a rather generic field theoretic effect governing the formation of high degeneracy macroscopic objects~\cite{Dvali:2020wqi} from featureless initial states.  In such a transition the object is formed in a maximally entangled superposition of all microstates~\cite{Dvali:2025wto}.  

 An extreme example is provided by a large BH formed in collision of two very high energy particles.   In order to compensate the exponential suppression, $e^{-S}$, of each microstate by their multiplicity~\cite{Dvali:2014ila,Dvali:2020wqi, Dvali:2022vzz}, the BH is formed in their fully entangled superposition.  We show that such a BH carries a maximal MB.

In contrast, for macroscopic sources, such as stars, a better knowledge of the memory encoding mechanism is required for determining $\mu$. 
Nevertheless, even without such details, we can establish certain boundaries. 
 We show that in case of BHs formed in a star collapse -- such as the ones observed in GW250114~\cite{LIGOScientific:2025rid,LIGOScientific:2025obp} -- the diversity of initial features can be sufficient for 
probing and constraining the encoding mechanism
already by the existing observational data.
 
 With these inputs, we give bounds on the critical exponent $p$ as  function of 
 $\mu$. The shifts of QNM frequencies in the ringdown due to SMB and resulting bounds on $p$ were already investigated closely in~\cite{Yuan:2025hls}; however, our formulas differ from the expressions used in that work, and we incorporate the theoretical uncertainties in $\mu$, significantly extending the estimates 
 of~\cite{Dvali:2025sog}.  This changes the constraints substantially. 

 Our analysis contains some additional novelties. 
In particular, in~\cite{Dvali:2025sog} only the shifts towards infrared were considered.  In this note, we focus on the ringdown and BH QNMs. We shall show that depending on model parameters the frequencies of QNMs can be shifted either towards the infrared or the ultraviolet. 

 The general message is that the sensitivity of $\mu$ towards the microscopic mechanism 
 of BH memory storage promotes the  classical GW astronomy to a powerful tool for probing this mechanism.

In section \ref{sec:review}, we briefly review previous results on MB. We then apply these findings to BH QNMs in section \ref{sec:QNM} to derive among others formula \eqref{masterFormula} for the frequency shift of QNMs. In section \ref{sec:estimatememoryload}, we estimate the memory load of astrophysical BHs and the resulting bound on the critical exponent $p$ that can be derived from GW250114. We discuss the conceptual difference between evaporation and QNMs in section \ref{sec:wayOut}, before we summarize our conclusions in section \ref{sec:conclusion}.

\section{Review}
\label{sec:review}

\subsection{Assisted gaplessness}
The essence of enhanced information storage and MB can be captured by the Hamiltonian~\cite{Dvali:2017ktv,Dvali:2017nis, Dvali:2018vvx,Dvali:2018xoc,Dvali:2018tqi,Dvali:2018xpy,Dvali:2018ytn,Dvali:2020wft}
\begin{equation}\label{H}
	\hat{H}_{\text{m}} = \epsilon_0 \hat{n}_0 + \sum_{k=1}^K \mathcal{E}_k(\hat{n}_0) \hat{n}_k \;.
\end{equation}
Here $\hat{a}_k^\dagger$ and $\hat{a}_k$, $k=0,\ldots, K$, are bosonic creation and annihilation operators that fulfill canonical commutation relations, $[\hat{a}_j,\hat{a}_k^{\dagger}] = \delta_{jk}$, $[\hat{a}_j,\hat{a}_k]  =   [\hat{a}_j^{\dagger},\hat{a}_k^{\dagger}] =0$, and form number operators $\hat{n}_k\equiv \hat{a}_k^{\dagger} \hat{a}_k$ with expectation values $n_k\equiv \braket{\hat{n}_k}$. 
 
 The modes are divided in two categories:  
the mode $\hat{n}_0$ with energy gap $\epsilon_0$ is the \textit{master mode}, and $\hat{n}_1$,$\dots$,$\hat{n}_K$ represent \textit{memory modes}.  The tasks between the two categories are split in the following way. The memory modes, which come in large diversity $K$, 
store information patterns in the sequence of their occupation numbers.
 These patterns are described as the states
 $\ket{n_1, n_2, \ldots, n_K}$ in the Fock space. 
 
  The task of the master mode is to create a local ``environment" 
  that minimizes the energy cost of the information pattern. 
   For achieving  this goal, it is 
crucial that the effective energy gaps $\mathcal{E}_k(n_0)$ of the memory modes depend on the master mode and have zeros for certain critical occupation numbers $n_0$.  Since we are mainly interested in physics near such points, the essence of the phenomenon is well captured by the following 
gap functions~\cite{Dvali:2017nis, Dvali:2018xpy,Dvali:2018ytn,Dvali:2020wft}:
\begin{equation} \label{effectiveGap}
	\mathcal{E}_k(n_0) = \left(1-\frac{n_0}{N_c} \right)^p \epsilon_k \;,
\end{equation}
where $\epsilon_k$ is the free energy gap of memory modes, $1/N_c$ sets the strength of an attractive interaction (with $N_c\gg 1$), and $p$ is an a priori undetermined critical exponent.

We must comment on the viability of the Hamiltonian (\ref{H}) with the 
gap function (\ref{effectiveGap}) as the right prototype for describing the BH. First, up to  inessential variations, one uniquely arrives at
this structure in describing any consistent quantum system with efficient information storage
\cite{Dvali:2017ktv,Dvali:2017nis, Dvali:2018vvx,Dvali:2018xoc,Dvali:2018tqi,Dvali:2018xpy,Dvali:2018ytn,Dvali:2020wft,Dvali:2024hsb}. Independently, it represents an effective description of the emergence of memory modes at 
criticality of the graviton coherent state describing the BH metric in a microscopic theory of quantum $N$-portrait~\cite{Dvali:2011aa, Dvali:2012en, Dvali:2013vxa, Dvali:2013eja, Dvali:2015ywa, Dvali:2015wca, Dvali:2016zqx}. 

 Indeed, it is easy to see how the above system provides the efficient information storage.
Once the master mode reaches a critical occupation $n_0 = N_c$, the gaps of all memory modes vanish, $\mathcal{E}_k(n_0)=0$. In this way, the master mode ``assists''~\cite{Dvali:2018tqi} the memory modes in becoming gapless.  Due to this, the states corresponding to their different occupation numbers $n_k$,
\begin{equation} \label{BHstate}
	\ket{\underbrace{N_c}_{n_0}, n_1, \ldots, n_K} \;,
\end{equation}
become degenerate in energy.  It is clear that the 
microstate degeneracy grows exponentially with the diversity $K$
of the memory modes.
 For example, even if the memory modes are restricted to qubits, $n_k=0,1$, they can form $2^K$ microstates and in this way account for an entropy $\sim K$.
The efficiency of information storage can be quantified by comparing the energy $\epsilon_0 N_c$ of the enhanced memory state \eqref{BHstate} with the would-be cost $E_P$ of the memory pattern in the absence of a master mode:
\begin{equation} \label{mu}
	\mu \equiv \frac{\epsilon_0 N_c}{p E_P} \;, \qquad 	E_P =   \sum_{k=1}^K \epsilon_k  n_k \;,
\end{equation}
where we remark that this definition of $\mu$ follows~\cite{Dvali:2025sog} while it differs from naming in previous papers (\eg~\cite{Dvali:2018ytn,Dvali:2020wft,Dvali:2025ktz}).

 We must note that, although the Hamiltonian 
(\ref{H}) is particle number conserving, it fully suffices for capturing the physical essence of the effect. This is due to the critical nature of the phenomenon. Correspondingly, the addition of all possible number-violating interactions gives no qualitative change and is unnecessary (see,~\cite{Dvali:2020wft, Dvali:2025sog}). 

However, the knowledge that 
the Hamiltonian gives an effective description of a bounded object 
originating  from a relativistic quantum field theory (QFT)
provides the following dictionary. Namely, for the object of radius 
$r_g$ saturating the 
QFT bound on the microstate degeneracy~\cite{Dvali:2020wqi}, the parameters achieve the values,
\begin{equation} \label{BHscalings}
	\epsilon_0 \sim r_g^{-1} \;, \qquad \epsilon_k \le  \sqrt{S} r_g^{-1} \;, \qquad  N_c\sim K \sim  S \;.
\end{equation}
Obviously, the above relations hold for a BH~\cite{Dvali:2018xpy, Dvali:2020wft, Dvali:2024hsb, Dvali:2025sog, Dvali:2025ktz}
with the following identification:  $r_g = 2 G M$ is the Schwarzschild radius and $S=2\pi M r_g$ corresponds to the Bekenstein-Hawking entropy.  We also notice that $\epsilon_0 n_0 \sim r_g^{-1} S \sim M$, \ie the
mass of the BH is carried by the occupation number of 
the master mode. This is natural since the composition of the classical gravitational field comes from the coherent state of the master mode~\cite{Dvali:2011aa, Dvali:2012en, Dvali:2013vxa, Dvali:2013eja, Dvali:2015ywa, Dvali:2015wca, Dvali:2016zqx}.
Additionally, the state \eqref{BHstate} is characterized by 
\begin{equation} \label{NmBound}
	N_P \equiv  \sum_{k=1}^K  n_k \lesssim S \;,
\end{equation}
but this bound does not need to be saturated~\cite{Dvali:2025sog}. It is possible that only a smaller number $N_P\ll S$ of memory modes is excited, and in this case their highest gap can be smaller, $\tilde{\epsilon}_k \sim \sqrt{N_P} \,r_g^{-1}$. For the maximal $N_P \sim S$, the efficiency parameter \eqref{mu} scales as 
\begin{equation} \label{MaxMB}
  \mu\sim \frac{1}{p\sqrt{S}}  \;.
\end{equation}
 In general, astrophysical BHs are expected to exhibit much higher $\mu$~\cite{Dvali:2025sog}.  The goal of the present paper is to provide a better understanding of this parameter.

\subsection{MB}

MB~\cite{Dvali:2018xpy,Dvali:2018ytn,Dvali:2020wft,Dvali:2024hsb} arises from the fact that the energy gaps \eqref{effectiveGap} of the memory modes only vanish for the critical occupation $n_0=N_c$. Once $n_0$ changes, the memory modes acquire gaps. This would lead to a huge energy cost. Due to this, the memory modes create a backreaction against any departure of the master mode from the critical value~\cite{Dvali:2018xpy,Dvali:2018ytn,Dvali:2020wft,Dvali:2024hsb}. In the Hamiltonian \eqref{H}, we can quantify this effect  in terms of~\cite{Dvali:2018ytn,Dvali:2020wft,Dvali:2024hsb, Dvali:2025ktz}
\begin{equation}
\label{eq:mutilde}
	\tilde{\mu} \equiv \sum_{k=1}^K n_k \frac{\partial  \mathcal{E}_k(n_0)}{\partial n_0} = -\frac{\epsilon_0}{\mu} \left(1-\frac{n_0}{N_c} \right)^{p-1} \,,
\end{equation}
which yields an effective contribution to the energy gap of the master mode 
\begin{equation} \label{effectiveGapMasterMode}
\mathcal{E}_0 \approx \epsilon_0+\tilde{\mu} = \epsilon_0\left(1- \mu^{-1} \left(1-\frac{n_0}{N_c} \right)^{p-1}\right) \;.
\end{equation}

MB has to set in at the latest when $|\tilde{\mu}|$ becomes of the order of the elementary gap $\epsilon_0$. This yields 
\begin{equation} \label{maximalDeltaN}
q\equiv	\frac{|\Delta n_0|}{N_c} \simeq \mu^{1/(p-1)} \;,
\end{equation}
where we denote the change of the occupation number of the master mode by
\begin{equation} \label{DeltaNAlpha}
	\Delta n_0 = N_c - n_0 \;.
\end{equation}
For maximal memory load, \ie the minimal value \eqref{MaxMB} of $\mu$,
\eq \eqref{maximalDeltaN} gives $q \sim (p^2 S)^{-1/(2(p-1))}$, in agreement with previous results~\cite{Dvali:2018ytn,Dvali:2020wft,Dvali:2024hsb,Dvali:2025ktz}.\footnote
{The case $p=2$, corresponding to $q\sim 1/{\sqrt{S}}$, receives additional motivation beyond MB~\cite{Dvali:2015wca,Michel:2023ydf}.}
For evaporating BHs, the quantity $q$ in \eqref{maximalDeltaN} corresponds to the mass fraction emitted semi-classically before the object is stabilized by MB. 

\begin{figure}
	\centering
	\includegraphics[width= 0.9\linewidth]{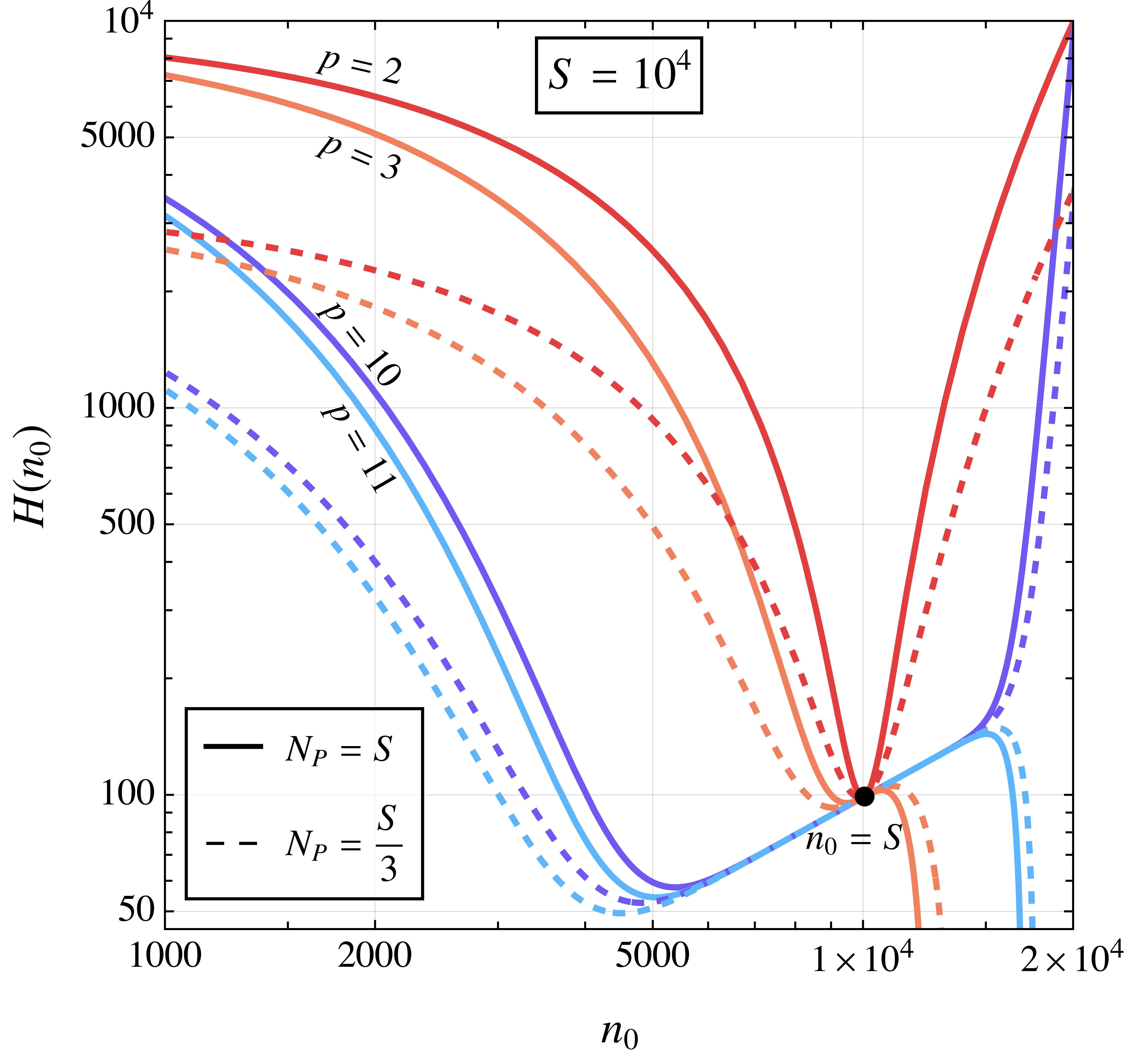}
	\caption{Hamiltonian for the master mode \eqref{H} with memory sector given by \eqref{effectiveGap}. Parameters as a function of entropy are given in \eqref{BHscalings} (for illustrative purposes, $\epsilon_k =\sqrt{S}\,r_g^{-1}$) and $N_P$ is defined in \eqref{NmBound}. As $n_0$ decreases, \ie the BH evaporates, $H(n_0)$ at first also decreases, in accordance with the semi-classical approximation. Eventually, however, a minimum is reached and $H(n_0)$ increases with decreasing $n_0$ -- this marks the onset of MB, which suppresses further decay.}
	\label{fig:effpot}
\end{figure}

\begin{figure}
	\centering
	\includegraphics[width= 0.9\linewidth]{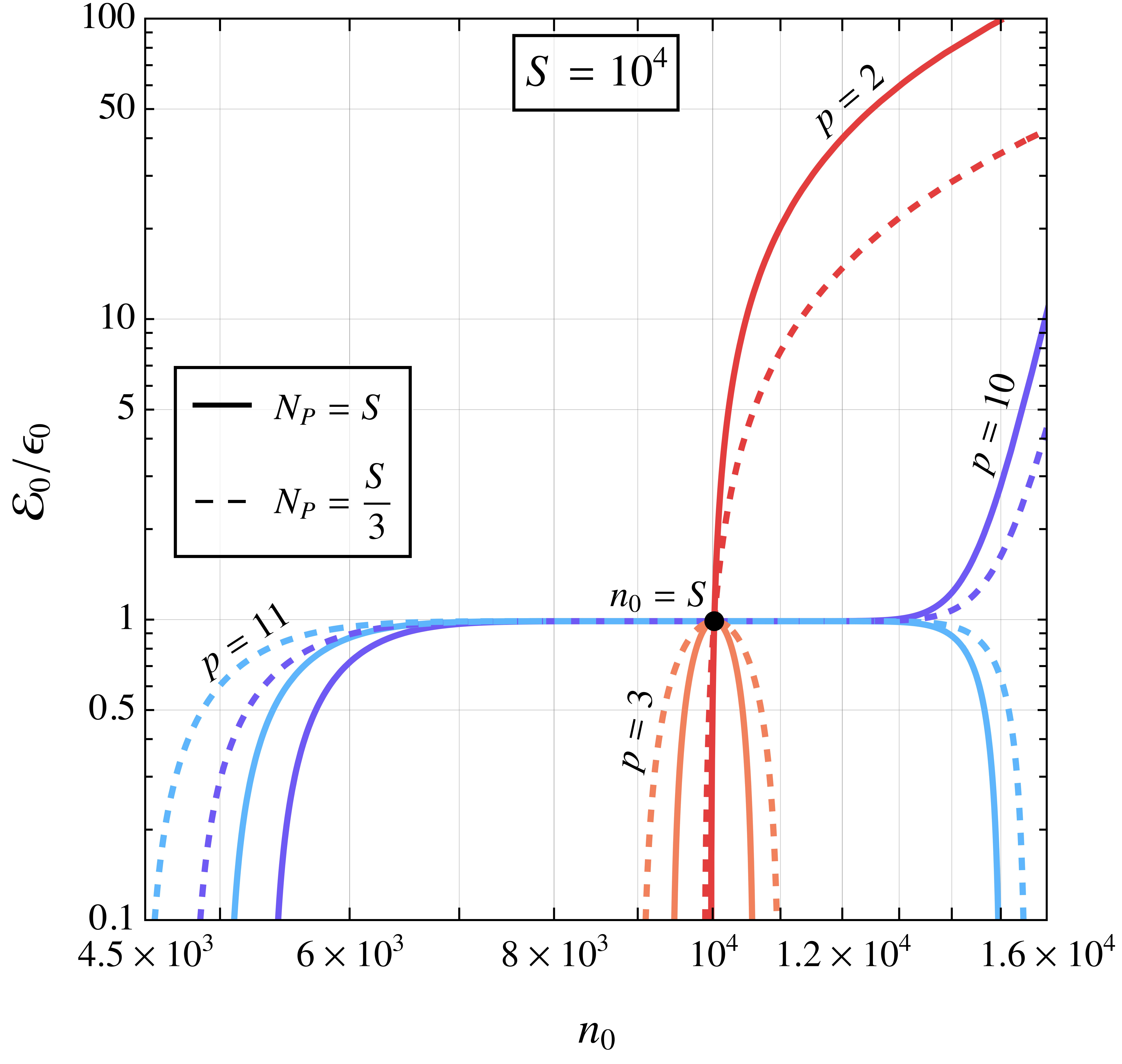}
	\caption{Effective gap of the master mode \eqref{effectiveGapMasterMode} for different values of the critical exponent $p$, defined in \eqref{effectiveGap}. Parameters as a function of entropy are given in \eqref{BHscalings} (for illustrative purposes, $\epsilon_k =\sqrt{S}\,r_g^{-1}$) and $N_P$ is defined in \eqref{NmBound}.}
	\label{fig:gap}
\end{figure}

The potential for the master mode as well as its effective energy gap are depicted in \figs \ref{fig:effpot} and \ref{fig:gap}, respectively. A semi-classical BH is achieved at the black dot $n_0 = S$. Moving away from that, the system is subjected to an energetic barrier characterized by the memory sector. Correspondingly, the effective energy gap $\mathcal{E}_0$ is macroscopically altered. The effect is governed by the critical exponent $p$ in \eqref{effectiveGap}.
For higher values of $p$, the potential is shallower and the onset of MB is further away from $n_0=S$. A similar effect is achieved if the memory load is lower as it can be seen from comparing the full and the dashed lines corresponding to $N_P = S$ and $N_P = S/3$ memory loads, respectively. We note that for odd values of $p$, lower energy values can be achieved as compared to the semi-classical case.\footnote{A seemingly unbounded from below energy for odd $p$ is not an issue since the Hamiltonian is an effective description around criticality.}

Importantly, the discussion thus far applies both to MB as a result of evaporation -- as was the focus of virtually all previous works~\cite{Dvali:2018xpy,Dvali:2018ytn,Dvali:2020wft,Dvali:2021bsy,Dvali:2021tez,Dvali:2023xfz,Dvali:2024hsb,Dvali:2025ktz,Contri:2025eod} -- and to SMB~\cite{Dvali:2025sog}, which is caused by an external perturbation. In particular, the critical relative change \eqref{maximalDeltaN} of the occupation number, beyond which MB has to set in, applies to both cases. The difference is that for evaporation, the timescale after which MB becomes visible is bounded from below by the speed at which the systems emits particles.
For a BH, this timescale is much larger than $r_g$. In contrast, SMB occurs on the timescale of perturbations, which is typically much shorter and on the order of $r_g$.

\section{Application to BH mergers}
\label{sec:QNM}

\subsection{BH scalings}

Following~\cite{Dvali:2025sog},  we shall now apply the SMB effect to BH mergers.  Although the process of merger is highly non-linear, the following physical reasoning allows to draw certain clean
 conclusions. The key point is that the classical evolution process can be viewed 
 as a sequence of coherent transitions between various modes. 
 In particular, the emission of the gravitational radiation during the ringdown phase represents a conversion of the perturbation of the master mode,
 describing a BH QNM, into a radiation mode. 
   At any given moment of time, the transition ``partner'' from 
   the radiation spectrum is chosen according to the resonant frequency matching the instantaneous frequency of the master mode.  This frequency 
   is of course set by the respective MB which changes in time as a function of the perturbation amplitude of the  master mode. 
   Correspondingly, the system continuously scans among the radiation modes searching for the transition partner of the resonant frequency.  
 This frequency is altered by the memory load, imprinting 
 the effect into the spectrum of the outgoing radiation
 (\ref{DeltaF1}).

Plugging in the scaling \eqref{BHscalings} corresponding to a BH, the deformed effective gap of the master mode, as derived in \eq \eqref{effectiveGapMasterMode}, becomes
\begin{equation} \label{effectiveGapMasterModeBH}
	\frac{\mathcal{E}_0}{\epsilon_0} \sim 1-\mu^{-1}\left(\frac{\Delta n_0}{S}\right)^{p-1} \;.
\end{equation}
Since the BH constantly scans the radiation modes to find a resonant partner, $\mathcal{E}_0$ sets the energy scale of emitted radiation~\cite{Dvali:2025sog}. Thus, the only remaining task is to relate $\Delta n_0$ to the perturbation of the BH metric.

An unperturbed BH can be represented as a coherent state $\ket{n_0}$~\cite{Dvali:2011aa, Dvali:2012rt, Dvali:2012en, Dvali:2013vxa, Dvali:2013eja, Dvali:2015ywa, Dvali:2015wca, Dvali:2016zqx}. Then, after the  perturbation induced by the merger, the state of the system corresponds to a deformation $\ket{n_0 + \Delta n_0}$ of the coherent state. Since the canonically normalized metric perturbation $\hat{h}$ can be expressed in momentum space as $\hat{h} \sim \hat{a}_0$ and $\hat{a}_0 \ket{n_0} \sim \sqrt{n_0} \ket{n_0}$, the deformation of the metric scales as
\begin{equation}
	|\delta h|^2  = \braket{n_0 + \Delta n_0|\hat{h}|n_0 + \Delta n_0}^2 - \braket{n_0|\hat{h}|n_0}^2 \sim  \frac{|\Delta n_0|}{r_g^2}  \;,
\end{equation}
where we took into account that the frequency of the master mode is $\epsilon_0 \sim r_g^{-1}$, \cf \eq \eqref{BHscalings}. Thus, the dimensionless metric perturbation $\delta g \sim \delta h / M_{\rm P}$ scales as 
\begin{equation} \label{deltaG}
	|\delta g|^2 \sim \frac{|\Delta n_0|}{S} \;,
\end{equation}
where $M_{\rm P}=1/\sqrt{G}$ is the Planck mass.
Plugging in the maximal value \eqref{maximalDeltaN}, at which MB has to set in, and using that $N_c\sim S$ for BHs, we arrive at the critical value of the metric perturbation
\begin{equation} \label{deltaGMax}
	|\delta g|^2 \sim \mu^{1/(p-1)} \,.
\end{equation}
Subsequently, we shall apply this formula to BH mergers and QNMs.

\subsection{BH QNMs}
For initial inspiral phase of a BH coalescence, we can identify $\hat{n}_0$ with the master mode of one of the initial BHs; see~\cite{Dvali:2025sog} for more details. Since energy is transferred to the new BH state and outgoing radiation, $n_0$ decreases and so drops below the relevant $N_c$. Thus, $\Delta n_0 > 0$ (see definition \eqref{DeltaNAlpha}) and $\mathcal{E}_0<\epsilon_0$. Consequently, \eq \eqref{effectiveGapMasterModeBH} shows that MB shifts the frequency of GWs towards the infrared~\cite{Dvali:2025sog}.

 However, the situation is different in ringdown. In this case, we identify $\hat{n}_0$ with the master mode of the final BH. Only at asympotically late times, after the emission of GWs corresponding to QNMs, the BH state $n_0 =N_c \sim S$ can be reached. In particular, this is necessary for arriving at the state of a ``young'' BH, not immediately subjected to MB, after the merger, as also assumed in~\cite{Zantedeschi:2024ram,Dondarini:2025ktz} (see moreover~\cite{Dvali:2021ofp,Dvali:2023qlk} for spin effects).
 
 In late stages of the ringdown, the BH can therefore be characterized by $n_0>N_c$ and so $\Delta n_0 < 0$. In this case, \eq~\eqref{effectiveGapMasterModeBH} makes evident that the direction in which the frequency of GWs is shifted depends on $p$. For odd $p$, the frequency of GWs is shifted towards the infrared. In contrast, for even $p$ the frequency increases, \ie is shifted towards the ultraviolet.  
 
In summary, plugging \eq \eqref{deltaG} into the effective gap \eqref{effectiveGapMasterModeBH} of the master mode and taking into account the shift toward the infrared or ultraviolet, a frequency $f_R$ is shifted to
\begin{equation} \label{masterFormula}
	f = f_R\left(1 - \mu^{-1} \left(-|\delta g|^{2}\right)^{p-1}\right)\;,
\end{equation}
as already summarized in \eq \eqref{DeltaF1}. Here we assumed $p$ to be integer and we note that $f\rightarrow f_R$ if either the memory load is small, $\mu \rightarrow \infty$, or if the amplitude of perturbations vanishes, $|\delta g|\rightarrow 0$.\footnote
{The formula used in~\cite{Yuan:2025hls}, which differs from \eq \eqref{masterFormula}, does not fulfill these two consistency conditions since a shift of the frequency that is independent of both $\mu$ and $|\delta g|$ was considered.}
Moreover, we point out that the frequency shift is expected to contain an additional mild dependence on the frequency~\cite{Dvali:2025sog}: $f = f_R\left(1 - \mu^{-1} \left(-f_R \,r_g|\delta g|^{2}\right)^{p-1}\right)$. At present, however, the relevant value of the metric perturbation $|\delta g|$ is not uniquely determined, and so we can absorb $f_R\, r_g$ in an appropriate redefinition of  $|\delta g|$.

This said, the presented arguments about the sign of $\Delta n_0$ must be taken with a grain of salt. In particular, the coherent state of gravitons describing the BH metric is a distribution also in the momentum space, dominated by the momentum $1/r_g$~\cite{Dvali:2011aa, Dvali:2012rt, Dvali:2012en, Dvali:2013vxa, Dvali:2013eja, Dvali:2015ywa, Dvali:2015wca, Dvali:2016zqx}. 
 At the same time, the memory modes, being angular momentum eigenstates,  have profiles in radial direction. Correspondingly, the perturbation 
 $\delta g$ that represents a distribution of perturbations of different master momentum modes, simultaneously climbs the walls of the gaplessness ``canyon" at different momentum
 locations. 
 As a consequence, the departure of a perturbation from the gaplessness point involves certain averaging in the MB space.
 
 Also, treating the perturbation in the ringdown phase as a small departure from the gaplessness point corresponding to the ground state of a newly formed BH implicitly assumes that the 
 field had sufficient time to restructure and settle around the new gaplessness point. Of course, it is conceivable that after the merger the BH does not -- or at least not immediately -- settle to the zero-burden ground state. In this case, the effect of SMB would be even more drastic and potentially persist after the ringdown phase. 

Overall, the discussed prototype model can certainly capture the essentials in terms of effective $\Delta n_0$ with the sign kept as a free parameter.

As is evident from \eq \eqref{masterFormula}, the effect of MB is small as long as $|\delta g|^{2(p-1)} \ll \mu$. Thus,
\begin{equation} \label{boundP}
	p \gtrsim 1 + \frac{\ln \mu}{\ln |\delta g|^2} \;,
\end{equation}
is sufficient for making the effect of MB smaller than the uncertainty of the experimental measurement. Since QNMs are only defined in the regime of linear gravity, we know that $|\delta g| \ll 1$. However, $|\delta g| \sim \mathcal{O}(10^{-1})$ can be sufficient for the suppression of non-linearities. This is consistent with the finding that in GW250114, QNMs can be identified before the observed strain has dropped by a factor of roughly $10$ as compared to the maximal amplitude~\cite{LIGOScientific:2025rid,LIGOScientific:2025obp}.

\subsection{Evolution after MB}

Eq.\@\xspace \eqref{deltaGMax} provides a bound on the amplitude of metric perturbation for which the MB effect becomes strong.  How the dynamics is affected beyond this point is a matter of a separate investigation. However, 
 certain useful phenomenological bounds can be derived even within the validity of our perturbative  treatment of the MB effect. Before doing so, some clarifying remarks are in order. 

 First, as in~\cite{Dvali:2025sog},  
 we assume that the MB effect cannot be avoided by a ``miraculous'' confinement of the evolution trajectory to a null MB surface. That is, we assume that the contribution of the  perturbation into the gap function is
 not compensated by a different master mode in such a way that the memory modes remain gapless.
For example, instead of \eq \eqref{effectiveGap}, we could consider a gap function that depends on a second 
master mode with occupation number $\tilde{n}_0$, 
\begin{equation} \label{effectiveGapSecondMasterMode}
	\mathcal{E}_k(n_0, \tilde{n}_0) = \left(1-\frac{n_0 + \kappa \, \tilde{n}_0}{N_c} \right)^p \epsilon_k \;,
\end{equation}
 where $\kappa$ is a relative coefficient. This structure allows for having perturbations that satisfy the constraint $n_0 + \kappa \, \tilde{n}_0 = N_c$ and therefore lie on a null MB surface. Obviously, any
such perturbation will respect the gaplessness of the memory modes,  
$\mathcal{E}_k(n_0, \tilde{n}_0)=0$, thereby resulting in a vanishing MB effect. 

An alternative version of a null MB trajectory can be realized if the master modes
control the gap functions of distinct sets of the memory modes. For example, 
	$\mathcal{E}_k(n_0) = (1-{n_0}/{N_c} )^p \epsilon_{k}$ and 	$\tilde{\mathcal{E}}_{\tilde{k}}(\tilde{n}_0) = (1-{\tilde{n}_0}/{\tilde{N}_c} )^{\tilde{p}} \epsilon_{\tilde{k}}$,
where, for generality, we allow all parameters, including the  
critical exponents $p, \tilde{p}$ and the critical occupation numbers $N_c, \tilde{N}_c$, to be different. Obviously, if the memory pattern is stored exclusively in the memory modes of the register $k$, the perturbation that only affects the master mode $\tilde{n}_0$ will correspond to a null MB trajectory.  

We dismiss the possibility of a null MB situation due to the following obvious reasons. First, generic initial perturbations, which can be chosen arbitrarily, do not respect it. Secondly, even if the null MB condition is fulfilled at some initial time, there is no (known) reason for the time evolution to preserve it. 

Of course, the MB effect will always try to minimize the gap. This means that any additional  master mode that was ``dormant'' in the initial state can be activated in order to minimize the gap function. But this already implies that the MB effect is operative and influences the dynamics, resulting in departures from the standard semi-classical picture that ignores the MB effect.  This is the phenomenon we are after. 

 Staying within the validity of our perturbative  treatment of the MB effect, we  arrive to two robust messages. 
 First, we identify the critical strength \eqref{deltaGMax} of metric perturbations above which the MB effect becomes strong. 
 Secondly, as in~\cite{Dvali:2025sog}, we predict the shift \eqref{masterFormula} of the frequency of QNMs, but allow for having both signs.  

\section{Estimate of memory load for astrophysical BHs}
\label{sec:estimatememoryload}

\subsection{The relation between the information content and the memory load}

 In general, the value of the MB parameter 
 $\mu$ for a given BH depends on the nature of the collapsing source that formed it.  If the source has high diversity of features, the minimal value of the 
memory load is  expected to be higher.  
Some estimates have already been made  in~\cite{Dvali:2025sog}. 
However,  diversity of the source only provides a lower bound as
the MB of a BH can strongly exceed the 
diversity count (and thus the information content) of the collapsing source. 
For example, a highly entangled 
state of all memory modes can carry a nearly maximal MB, despite a very small information content. This is due to a strong correlation of memory modes in such a state which leads to the ``shuffling'' of information~\cite{Dvali:2025sog} as we shall now discuss.

In order to prepare a basis, we 
first point out the following universal QFT effect characteristic of objects of high macrostate degeneracy~\cite{Dvali:2020wqi}. When formed in a transition 
from a featureless initial state, the object
materializes in a superposition of all microstates~\cite{Dvali:2025wto}. The reason is that such a superposition maximizes the transition probability,  making up for the exponential suppression of individual microstates. This is especially clear when the degeneracy of the microstates is due to a symmetry, as we shall discuss in the next section.

 The extreme example is given by a  BH 
 formed in a two-particle collision.
 Already the classical General Relativity suggests that a  BH must form as long as the  center of mass 
 energy of the colliding quanta is above the BH mass
 and the impact parameter is less than its gravitational radius.   Due to this physical argument, the BH formation in high energy collision has been discussed for some time~\cite{tHooft:1987vrq, Amati:1987wq, Gross:1987kza}.  
Our goal is to match this general understanding, which is based on the semi-classical picture, with the explicit computations in the microscopic quantum theory~\cite{Dvali:2014ila, Addazi:2016ksu}, in order to estimate the resulting memory load.

  Consider, for example, the formation of a BH 
  in $s$-wave collision of two spin-$0$ bosons of center of mass energy $M$.  
  Of course, we assume that $M \gg M_{\rm P}$, so that 
  the formed BH is macroscopic.    
  Although the initial state is almost featureless, the resulting BH is expected to carry 
  close to the maximal memory load \eqref{MaxMB}.
   This follows from the matching of semi-classical 
   and quantum pictures. 
    
    The semi-classical arguments unambiguously tell us that 
    a BH must form, with close to unit probability,  when the 
  particles become localized within the Schwarzschild radius of 
  the center of mass energy.  
  This must be matched by the quantum picture.

 Now, in general, the formation of a non-perturbative object in a two-particle 
  collision process is expected to be exponentially suppressed. This suppression is a generic feature of QFT and is not specific to gravity. The detailed explanation of this phenomenon  as well as the relevant references can be found in~\cite{Dvali:2022vzz}.  Here we shall briefly recount the main point, which is relevant for our discussion. The catch is that a non-perturbative process in which a macroscopic object, say a soliton or a BH, is formed in a two-particle collision, represents a multi-particle process of the type, 
  \begin{equation} \label{2toN}
      2 \rightarrow N \,,
  \end{equation}
in which the two initial quanta of the center of mass energy $E$ transit into a large number, $N\gg 1$,  of the soft ones.  The quanta are soft because of the division of the initial energy among them.  
These $N$ soft quanta are the constituents of the coherent state describing the resulting macroscopic object.  It is this multi-particle nature of the process that makes it highly suppressed.  

  While the quantitative factors of the suppression can differ from case to case, its exponential nature is universal. In fact, there exists a general proof that the rate of the transition process of the type (\ref{2toN}) in QFT is bounded from above by $e^{-N}$~\cite{Dvali:2020wqi}.   
     
     In case of a BH formation, the exponential suppression is confirmed by the explicit calculation~\cite{Dvali:2014ila, Addazi:2016ksu} in which the BH's classical metric is resolved as the coherent state 
   of $N$-gravitons of energies $\sim 1/r_g$ and the mean occupation number $N =S$~\cite{Dvali:2011aa}. 
    In this picture, the BH formation in a two-particle collision is described as a transition of the type (\ref{2toN}), with $N=S$ final gravitons   representing the master mode of the BH.  
    The calculation of~\cite{Dvali:2014ila, Addazi:2016ksu}
    shows that the formation of each particular microstate $\ket{{\rm BH}}_{\rm mcrst}$ of the BH in a two-particle scattering is indeed suppressed as\footnote{ Interestingly, this suppression matches the naive Boltzmann factor $\exp(-M/T)$ of the inverse transition process in which a BH of Hawking temperature 
    $T=1/R$ would decay in only two quanta.
    Notice that, since $M/T \sim S$, the Boltzmann  factor gives a suppression of the type obtained by the explicit 
    calculation of~\cite{Dvali:2014ila}.  
    Of course, this coincidence can at best be treated as indicative, since the thermal approximation is invalid in such a process. However, it can still be used as a qualitative argument in support of the exponential suppression of the process.
    }
\begin{equation} \label{RateS}
    \Gamma_{2 \rightarrow \rm BH_{mcrst}} \sim {\rm e}^{-S}\, .
\end{equation} 
      Now, as discussed in~\cite{Dvali:2014ila}, 
      in order to match the (semi)classical picture of BH formation in two-particle collision process, the 
      suppression $e^{-S}$ must be compensated by the number 
      of microstates 
      \begin{equation}
      \label{nstes}
      n_{\rm st} = e^S\,.
      \end{equation}
      Correspondingly, the 
      outcome is that, although the production of any given microstate is suppressed as (\ref{RateS}), due to their exponential diversity, 
      the BH is formed with order-one probability. 
  As discussed in~\cite{Dvali:2020wqi}, this counting universally applies to a two-particle scattering production of a generic macroscopic object that saturates the QFT bound on the microstate degeneracy.  

  In order to make the above more clear,  let us label the BH microstates 
  $\ket{\rm BH}_r$ with a single index $r =1,2,...,n_{\rm st}$, where $
  n_{\rm st} = e^S$ (see \eq \eqref{nstes}).
  In this labeling,  each value of the index $r$ refers to a specific  excitation pattern (\ref{BHstate}) of the memory modes.  We are looking  for a  BH formation starting at some initial time 
  $t_{i}$ from an initial two-particle state 
  $\ket{2}$ of center of mass energy $M$. 
  This state is evolved to a BH state at some final time $t_f$ by an unitary time-evolution  operator $U(t_{i}, t_{f})$. 

   The transition matrix element to each microstate is exponentially suppressed~\cite{Dvali:2014ila},
\begin{equation} \label{2BHr}
 |\bra{\rm BH_r} U(t_{i}, t_{f}) \ket{2}| = \ex^{-\frac{S}{2}} \,.
\end{equation} 
   Correspondingly, the right final state to which an unsuppressed transition takes place is described by the following superposition 
 of all microstates, 
 \begin{equation} \label{BHf}
 \ket{{\rm BH}_f} = \ex^{-\frac{S}{2}} \sum_{r =1}^{n_{\rm st}} c_r\ket{{\rm BH}_r}  \,,
\end{equation} 
   where $c_r$-s are the phase factors that take care of potential  destructive interference between various matrix elements. It is then clear that 
\begin{align} 
 &|\bra{{\rm BH}_f} U(t_{i}, t_{f}) \ket{2}| \nonumber\\
 & \qquad = \ex^{-\frac{S}{2}} |\sum_{r =1}^{n_{\rm st}} c_r\bra{{\rm BH}_r}  U(t_{i}, t_{f}) \ket{2}| \sim 1 \,.\label{2BHf}
\end{align} 
So far, the above reproduces the  results of~\cite{Dvali:2014ila} and~\cite{Dvali:2020wqi}
of understanding the unsuppressed production 
in two-particle scattering  
of a BH or other objects of maximal microstate entropy. In this respect, we must note that the property of materialization in a superposition over the degenerate microstates is not exclusive to BHs and is universally shared by solitonic objects with high microstate entropy, such as vacuum bubbles materialized in a tunneling or a collision process~\cite{Dvali:2025wto}.

 Now, this has important implication for the
 BH MB effect. Since in \eq (\ref{BHf}) all the microstates are equally represented,  it is clear that the expectation values of all memory modes over this state are equal, 
 \begin{equation} \label{VEVn}
 n_j = \bra{{\rm BH}_f}\hat{n}_j \ket{{\rm BH}_f} \sim 1 \,.
\end{equation} 
  For example, for qubits the above expectation value is $n_j = 1/2$.

   Thus, we reach the following important conclusion. The state in which the BH is formed in a two-particle collision is expected to carry a close to a maximal memory load, \ie the resulting $\mu$-parameter saturates the bound \eqref{MaxMB}.\footnote{From this discussion, 
    we can conclude that a PBH formed  via the early-confinement mechanism of~\cite{Matsuda:2005ey,Dvali:2021byy,Z26toappear} carries a maximal memory load \eqref{MaxMB}.  In this scenario, the PBH is formed in the collision of two quarks that are accelerated to macroscopic energies via the confining string.}

 Thus, the memory load can largely exceed the
 information  content of the system.  Indeed, as 
 discussed in~\cite{Dvali:2025sog}, in the state of the type 
 (\ref{BHf}) the memory modes are highly entangled, and the information carried by them 
 is negligible, \ie information is ``shuffled''. 
 Nevertheless, the memory load is close to maximal.   This distinction  between the memory load and the corresponding information must be 
 kept in mind for understanding $\mu$.  
    
\subsection{The role of symmetries in correlating memory modes}

 Let us now discuss the case in which the formation of a macroscopic object 
 in the superposition of microstates~(\ref{BHf}) is enforced by a symmetry~\cite{Dvali:2025wto}.   
     In case of a BH, of course, some of the microstates obey the selection rules by a symmetry. 
       In particular,  they ought to preserve the angular momentum. Such states must be formed in a symmetric 
 superposition matching the conserved quantum numbers of the 
 initial state.
For example, in $s$-wave collision the total state of the memory modes  must carry zero spin.

 In general, under any symmetry group of the Hamiltonian 
 also respected by the initial state, the memory modes 
 must form an invariant superposition  of the   type~(\ref{BHf})~\cite{Dvali:2025wto}.  
 For example, let us assume that initial particles forming a BH transform  as fundamental representation 
 under some exact symmetry $O(N)$. 
   According to the fundamental rule of BH informatics, 
every single species of a free quantum has a counterpart 
  gapless memory mode in the BH spectrum~\cite{Dvali:2025sog}.  In particular, this explains
  the independence of the BH's Bekenstein-Hawking 
  entropy on the number of particle species, $N_{\rm sp}$. The enhancement of the number of memory modes by $N_{\rm sp}$ exactly compensates the effect of lowering of the gravitational cutoff 
  to the species scale $\Lambda_{\rm sp} \equiv M_{\rm P}/\sqrt{N_{\rm sp}}$~\cite{Dvali:2007hz, Dvali:2007wp, Dvali:2008fd, Dvali:2008ec, Dvali:2009ks, Dvali:2010vm}. 
  Correspondingly, the total number of the gapless memory modes, 
  which determines the BH entropy, 
  scales as $(r_g\Lambda_{\rm sp})^2N_{\rm sp} = 
  M_{\rm P}^2 r_g^2$ and is independent  of $N_{\rm sp}$~\cite{Dvali:2025sog}. 
 
  In the present example, the memory modes
 coming from additional species transform as 
  fundamental representation of the $O(N)$-group, 
$\hat{a}_{\alpha}(k)$, where $\alpha =1,...,N$  and 
$k$ stands for other characteristics such as the momentum. 

Let the initial scattering quanta that form a BH carry a definite $O(N)$-index, $\alpha =1$. 
  The initial state is thus invariant under the $O(N-1)$ subgroup. 
  This symmetry must be inherited by the BH state. 
   Notice that this by no means precludes the 
  memory modes  with $\alpha \neq 1$ from being occupied. 
  Rather, the only constraint is that they must be occupied 
  in a way to form a $O(N-1)$-invariant state. 
    Taking $\ket{0}$ to be a BH vacuum in form of the empty memory pattern, the  $O(N-1)$-invariance forces 
    the BH state to be a superposition of the invariant states of the form~\cite{Dvali:2025wto} 
 \begin{align} \label{NCstates} 
\left(\sum_{\alpha = 2}^{N}\hat{a}_{\alpha}^{\dagger}\hat{a}_{\alpha}^{\dagger}\right)^{n} \ket{0} , 
\end{align} 
where $n$ is an integer. 

   Notice that 
   the occupation of $\alpha \neq 1$ memory modes 
  is not only possible but is in fact mandatory. This is because such states 
  contribute into the microstate degeneracy factor which is 
  necessary for  compensating the  exponential suppression of multi-particle amplitudes responsible for BH formation.

Now, if the collapsing source is a multi-particle state, 
 the count is somewhat different since the transition
 between two classical states may not be exponentially suppressed even if the microstate degeneracy is low. 
  To our knowledge, an explicit  evaluation of such a collapse as of a multi-graviton process 
  in the spirit of~\cite{Dvali:2014ila}, \cite{Dvali:2020wqi, Dvali:2022vzz}
  and its matching to 
  a classical collapse has never been 
  performed. We shall therefore refrain from making  strong assumptions in such cases and will solely rely on relatively well-understood factors.  
    In particular, we shall keep in mind that when a BH is produced, its memory load can exceed the diversity matching of the collapsing source, as we shall now discuss in detail.

\subsection{Qualitative picture}

Having outlined the above key concepts, we would like to elaborate on the connection between the diversity of the  features of the source and the MB parameter $\mu$.  
 The quantities $E_P$ and $N_P$ are uniquely defined  in terms of the excited memory modes of the storer (\ie a BH) and their energy gaps $\epsilon_k$. 

The key question is how these quantities are determined by the parameters of the source. In order to understand this, we must introduce the following two notions. \begin{enumerate}
    \item The first one is the diversity of the \textit{uncorrelated} features of the source, which we shall denote by 
$\tilde{N}_P$. This number is basis-independent, \ie it is independent of the choice and labeling of the degrees of freedom describing the source.  
\item The second quantity is the total number of features. This can be much larger than $\tilde{N}_P$ if there  exist correlations among the features of the source.  This number is basis-dependent in the sense that it depends on the number and the diversity of degrees of freedom that are used to label the state of the source. 
\end{enumerate}

 After the source collapses, the features are encoded in the pattern of BH memory modes of length $N_P$ and levels $\epsilon_k$.  This determines $E_P$ and correspondingly $\mu$.  However, the way $\tilde{N}_P$ translates into 
 $N_P$ depends on the microscopic encoding mechanism. 
 It is very important to understand that the number of occupied memory modes does not need to be minimal, \ie it is possible that $N_P \gg \tilde{N}_P$.  
 
 Let us explain this on a specific example.  We fix the basis 
 in which the state of the source $\ket{n_1',n_2',...,n_{N_s}'}$
 is described by $N_s$ degrees of freedom with occupation numbers $n_j', ~ j =1,2,...,N_s$.  These are the degrees of freedom describing the source and should not be confused with the BH memory modes. In particular, $n_j'$-s  need not be gapless. 
 
 Let us assume that the numbers $n_j'$ satisfy $N_s-1$ constraints. Thus, in this state, the number of uncorrelated features is equals to one, $\tilde{N}_P =1$.  When the source collapses, the information about the state is encoded in BH memory modes. A naive expectation is that, due to the lack of uncorrelated features, the encoding takes up a very small fraction of the memory space, \ie the number of memory modes involved can be negligibly small. However, this number depends on the specifics of the encoding mechanism and cannot be determined from first principles.

In order to understand this, let us imagine that due to the specifics of the  microscopic mechanism, this pattern is copied into the BH memory modes with one-to-one correspondence. The resulting BH memory pattern $\ket{n_1,n_2,...,n_{N_P}}$, will correspond 
 to $N_P = N_s$.  Thus, the number of memory modes involved in the pattern can be arbitrarily larger than $\tilde{N}_P$. 
 
 Of course, the copying mechanism can be even less efficient, resulting into $N_P \gg N_s$.  The bottom line of this discussion is that the number of excited memory modes {\it a priori} can be  much larger  than the number 
 $\tilde{N}_P$ counting the uncorrelated features of the source. 
 
 Additionally, depending on the specifics of the encoding mechanism, the gap of the memory modes can exceed the minimal one,  $\epsilon_k > \sqrt{N_P} \,\epsilon_0$. In summary, this gives the interval
 \begin{equation} \label{Interval1}
     \tilde{N}_P^{3/2} \lesssim \frac{E_P}{\epsilon_0} \lesssim S^{3/2} \;,
 \end{equation}
where the left/right bound corresponds to maximal efficiency/inefficiency.  As an example, inefficiency can be enforced by the presence of a symmetry, as previously discussed -- \cf \eq \eqref{NCstates}.

\subsection{Estimates of the memory load}
 In order to illustrate the sensitivity of $\mu$ towards the encoding mechanism, we shall provide extreme estimates of 
 $N_P$ in which the BH memory pattern inherits either all 
 diverse features of the source or only the number $\tilde{N}_P$
 of uncorrelated ones.  
 
  In the first  extreme regime, we shall essentially repeat the analysis of~\cite{Dvali:2025sog} for two examples: 
  {\it 1)} a collapsing thermal ball of radiation; and {\it 2)} 
  a collapsing star. First, we shall assume that the pattern is copied to BH memory modes fully, without reducing their number by the correlations in the source. 
  Next, we repeat
  both examples by narrowing the copied pattern only to uncorrelated features $\tilde{N}_P$. 
 Using these results, we estimate $\mu$ for all cases.  
  
  For our estimates, we shall assume that the features are encoded into the register of the memory modes with the lowest possible $\epsilon_k$ compatible with a given $N_P$. This yields
   \begin{equation}
      \epsilon_k = \sqrt{N_P} r_g^{-1} \,, 
    \end{equation}
  and correspondingly,  
  $E_P\sim r_g^{-1} N_P^{3/2}$. 
Then we get from \eq \eqref{mu}~\cite{Dvali:2025sog}, 
 \begin{equation}
 \label{eq:mudef}
 	\mu \sim \frac{S}{N_P^{3/2}} \,,
 \end{equation} 
 where we left out the power-law dependence on $p$.
  The key question is the estimate of $N_P$.

\subsubsection{$\mu$ for a radiation ball}  

 First, we consider
 a sphere of radius $R$ filled with a single 
 species of temperature $T \gg 1/R$. The number of 
 excited degrees of freedom scales as
 $N_{\rm in} \sim (TR)^3$.  If BH memory modes inherit all the features of the state, regardless of their correlations, the number involved in the pattern would be  $N_P  \sim N_{\rm in}$. Using $M\sim T (TR)^3$, we then get from \eq \eqref{eq:mudef}:
  \begin{equation} \label{muofRAD}
    \mu \sim (RT)^{3/2} \frac{T^2}{M_{\rm P}^2} \,.
\end{equation}
 For example, for  $T\sim\, \text{keV}$ and $R\sim 10^{12}\,\text{cm}\sim 10^{26}\,\text{GeV}^{-1}$, we have $\mu \sim 10^{-20}$.

Let us now show how different $\mu$  would be in the opposite extreme of the encoding mechanism in which 
the BH memory pattern only employs the number of memory 
modes equal to $N_P = \tilde{N}_P$.
   The number of uncorrelated features $\tilde{N}_P$ 
   is determined by factors that take the system out of an exact thermal equilibrium. More specifically, such departures are accounted for by the following dimensionless quantities that 
   depend of time and space gradients of the temperature~\cite{Dvali:2015aja}, 
   \begin{equation} \label{XiS}
 |\dot{T}|/T^2\,,~~ |\vec{\nabla} T|/T^2 \,.      
   \end{equation} 
One would say that, $\tilde{N}_P$ is suppressed relative to $N_{\rm in}$ by the largest of the above factors.  

Let us make some estimates. If the radius changes in time and the evolution is approximately adiabatic, we have 
$RT = {\rm const.}$. Taking the time derivative, we get 
 $\dot{T}/T^2 = -\dot{R}/(RT) = -\dot{R} N_{\rm in}^{-1/3}$, 
which for $\dot{R}= -1$ yields
$N_P =|\dot{T}/T| N_{\rm in} = N_{\rm in}^{2/3}=(RT)^2$. In contrast with 
\eq \eqref{muofRAD}, this gives, 
\begin{equation} \label{muRADCorr}
    \mu = (RT)^3 \frac{T^2}{M_{\rm P}^2} \,.
\end{equation}
 For  example, for  $T\sim\, \text{keV}$ and $R\sim 10^{12}\,\text{cm}$,
 we get $\mu \sim 10^{10}$.

 The above estimates can be applied to PBHs formed in the radiation dominated era by the collapse of the Hubble patch, in which case $R$ must be taken as the Hubble radius, $R_{\rm H} \sim M_{\rm P}/T^2$. However, as discussed in~\cite{Dvali:2025sog}, there is a caveat which indicates that Hubble-sized PBHs can carry a maximal memory load \eqref{MaxMB} regardless of the features contributed by the radiation. The reason is that the dominant diversity of the Hubble patch comes from the gravitational memory modes that carry Gibbons-Hawking entropy~\cite{Gibbons:1977mu}.  This entropy scales as $S_{\rm GH} \sim (M_{\rm P} R_{\rm H})^2$. 

Although, traditionally, the Gibbons-Hawking entropy has been discussed for de Sitter like states, there is no reason for disregarding it in case of other cosmological space-times. 
Of course, for a radiation-dominated Universe of temperature $T \ll M_{\rm P}$, the corresponding Gibbons-Hawking temperature $T_{\rm GH}$ is completely negligible as compared to the temperature of radiation, $T_{\rm GH} \sim T \, (T/M_{\rm P})$. In contrast, the Gibbons-Hawking entropy fully dominates over the entropy of radiation, 
$S_{\rm GH} \sim S_{\rm rad} \, (M_{\rm P}/T) \gg S_{\rm rad}$. 
 
 Correspondingly,  the maximal diversity of features that can be carried by the Hubble patch could come from the Hubble memory modes rather than from the radiation source. 
 
 The possibility of the MB effect due to Hubble memory modes has been considered previously in the context of de Sitter and inflationary universes~\cite{Dvali:2018ytn, Dvali:2021bsy}.  
  Our discussion is not about 
 generalizing this effect  to 
  radiation or matter dominated cosmologies.  The only important information we extract from the previous analysis is that the Hubble memory modes can be a dominant source of diversity for the
 PBHs that form on the Hubble scale.  
 Such PBHs therefore can inherit the maximal memory load  \eqref{MaxMB} from the Hubble memory modes~\cite{Dvali:2025sog}. 

 On the other hand, for BHs formed by the collapse of radiation spheres 
of much smaller radii, $R \ll R_{\rm H}$, the Hubble memory modes can be safely ignored. For such BHs the presented estimate of $\mu$ purely from the thermal bath is expected to be legitimate.

\subsubsection{$\mu$ for a collapsing star}

 Let us now move to estimating $\mu$ from a 
 collapsing star. In particular, we wish to estimate $\mu$ for GW250114. If BHs with $M \sim 30\, M_\odot \sim 10^{58}\,\text{GeV}$ originate from the core collapse of O-type main-sequence stars, one can roughly expect $R \sim 10\, R_\odot \sim 10^{12}\,\text{cm}$ (see~\cite{Holgado2025}) and $T\sim M R_\odot/(R M_\odot)\,  T_\odot \sim 10^{-6} \,\text{GeV}$ (see~\cite{Kippenhahn:2012qhp}). Then the number of photons is $N_{\gamma}  \sim (RT)^3 \sim 10^{60}$ and the number of baryons (ions) is on the same order ($N_i \sim M/\,\text{GeV}\sim 10^{58}$).\footnote
{For a non-relativistic species of mass $m$, the highest reachable momentum is $k_{\text{max}}\sim \sqrt{m T}$, which is larger than the corresponding $k_{\text{max}}\sim T$ of a relativistic species. However, the highest momentum level for which all modes can be occupied simultaneously is bounded by the number of available non-relativistic particles. Therefore, the contribution of baryons to $N_P$ cannot exceed the number of baryons.}

 First, we estimate the total number of features, ignoring the correlations, and assume 
 the encoding mechanism that copies them.   Using photons for the estimate, we then get from \eq \eqref{eq:mudef},
\begin{equation}
\mu \sim \frac{S}{(TR)^{9/2}} \;,
\end{equation}
or equivalently
\begin{equation} \label{masterFormula2}
	\mu \sim \frac{M^2}{M_{\rm P}^2 (TR)^{9/2}} \;.
\end{equation} 
Plugging in the numbers for the star, \eq \eqref{masterFormula2} yields
\begin{equation} \label{muEstimate}
\mu \sim 10^{-12} \;.
\end{equation}

As the opposite extreme, let us now assume that the encoding mechanism only 
 uses the number of memory modes given by uncorrelated 
 features $\tilde{N}_P$.  Relative to the occupation number of the source, this quantity is sharply reduced by correlations. For illustrative purposes, we shall consider the following maximal estimate of correlations.
 
We will assume that the only out-of-equilibrium feature of an old star is its space-dependent chemical composition. To derive to what extent unequal numbers of elements in different places are preserved, we compute the typical displacement due to diffusion,
\begin{equation}
    x \sim \sqrt{l v t} \sim \frac{T}{\alpha} \sqrt{\frac{v t R^3}{N_i}} \;,
\end{equation}
where $v$ is the typical velocity and we estimated the mean free path $l\sim R^3/(N_i \sigma)$ via Coulomb scattering. Therefore, the number of correlated ions is
\begin{equation}
    n_c \sim \frac{N_i x^3}{n_{\text{sp}} R^3} \sim \frac{v^{3/2} T^3 R^{3/2} t_\star^{3/2}}{n_{\text{sp}} \alpha^3 N_i^{1/2}} \left(\frac{t}{t_\star}\right)^{3/2} \;,
\end{equation}
where we take the lifetime of the star on the order of $t_\star \sim  10^7 \, \text{y} \sim 10^{38} \, \text{GeV}^{-1}$ and $n_{\text{sp}}$ accounts for the number of different types of ions with sizable abundance. Using $v\sim 10^{-3}$ as the typical velocity of protons at $T \sim 10^{-6} \,\text{GeV}$ and $n_{\text{sp}}\sim O(10)$ , we arrive at 
\begin{equation}
    n_c \sim  10^{50} \left(\frac{t}{t_\star}\right)^{3/2} \;.
\end{equation}
This gives $N_P\sim N_i/n_c \sim 10^{8}$, \ie in this estimate only $\sim 10^{8}$ features remain uncorrelated. Thus, instead of \eq \eqref{masterFormula2}, we get
\begin{equation}
    \mu \sim n_c^{3/2} \frac{M^2}{M_{\rm P}^2 N_i^{3/2}} \;,
\end{equation}
which yields 
\begin{equation}
    \mu \sim 10^{66} \,.
\end{equation}

We can now come back to constraining $p$ via  \eq \eqref{boundP} from the observation of QNMs in BH mergers. For BHs with $\mu^{-1}\gg 1$, we would get a strong signal for $O(1)$ values of $p$. For example, a large $\mu^{-1}$ would generically follow if memory modes were occupied uniformly, independently of $\epsilon_k$. In contrast, the effect of SMB on BH QNMs would remain small for $\mu^{-1} \ll 1$. Such BHs would yield GWs that -- at least at the current level of observational precision -- are indistinguishable from the semi-classical prediction.
 
 The lesson we learn is that, without understanding of the microscopic encoding mechanism, the knowledge of the source is  not sufficient for determining $\mu$ of the resulting BH. 
 Depending on this mechanism, for one and the same source, $\mu$ can take values in a large interval,
 ranging from direct observational prospects in BH mergers to no readable signal at all. In this situation, we take the attitude in which \eq \eqref{boundP} is treated as
 the phenomenological constraint on $\mu$ and $p$.
 Needless to say, in each regime of the encoding mechanism, 
 a phenomenological study of MB in BH QNMs 
 -- \eg along the lines of~\cite{Yuan:2025hls} -- remains to be performed.

\section{Can MB be avoided?}
\label{sec:wayOut}

The MB effect~\cite{Dvali:2018xpy,Dvali:2018ytn,Dvali:2020wft,
Alexandre:2024nuo, Dvali:2024hsb}
implies that the information carried by a BH backreacts on any sort of perturbation that is taking it
away from its ground state. In particular, this happens during BH evaporation. 
 The MB effect goes against the 
standard treatment in which the BH evaporation is assumed to be self-similar by extrapolating the semi-classical regime
over the BH's entire life. 
We shall refer to this treatment as the 
 ``simplistic picture''.

 We must note that even before the discovery of the MB effect, it has been understood~\cite{Dvali:2015aja} that the simplistic picture is invalidated by 
 $\sim 1/S$ corrections to thermality, which are inevitable due to the change of the BH temperature, $\dot{T}/T^2 \sim 1/S$.  However, the 
 MB effect makes the inconsistency of the standard treatment very explicit. 
Indeed, if the BH memory load is significant, 
there is no logical way in which the simplistic
picture can be accommodated.  On one hand, at the initial stages of evaporation, the information stored in memory modes is preserved by a BH, and on the other hand, the memory space shrinks. The MB backreaction is therefore unavoidable. 

 The MB effect is expected to be also present during any other departure of the BH from its ground state, in particular for QNMs excited during BH mergers~\cite{Dvali:2025sog}. In this case, the MB effect is swift and takes place on the timescale of the merger. We have seen that already within the validity of our perturbative treatment, observational data appears to be capable of probing the BH MB parameters.  This is both exciting and thought provoking. It makes one wonder whether, unlike the evaporation case, from which there is no way out,  the validity of the simplistic picture could be saved for the merger dynamics. 

 The difference with the evaporation case is that in mergers the absolute volume of the memory space grows.  It is conceivable that this leaves room for a possibility of the merger dynamics taking place entirely on a null MB surface, \ie a trajectory on which the gaps of the memory modes vanish. 
 
 Of course, the MB effect always tries to confine the system to such a surface. However, irrelevance of MB would require that this evolution takes place regardless of MB 
 rather than being imposed by it. 
 We could not come up with any consistent QFT logic 
 that would justify such a miracle. However, even
 if the miracle can be justified, \eg by a hidden symmetry, this would reveal a completely new law governing BH physics. Namely, the law would imply that regardless of the \textit{quantum} parameter $\mu^{-1}$, \textit{classically}, all BHs evolve on the trajectory that respects it. 
 
  Again, we stress that even if BH mergers were subjected to such a miracle, avoiding the MB effect during the evaporation process is impossible due to the inevitable shrinkage of the memory space.

Another important question is whether the MB carried by a BH can be reduced by external perturbations. This discussion shall be given in~\cite{DGKR}.  
  Of course, regardless, the analysis and the results of the present paper remain fully intact with respect to the effective $\mu$.

\section{Conclusion}
\label{sec:conclusion}

The robustness of the \emph{memory burden (MB)} effect~\cite{Dvali:2018xpy,Dvali:2018ytn,Dvali:2020wft,
Dvali:2021tez, Alexandre:2024nuo, Dvali:2024hsb} lies in the fact that it emerges from well-established properties of BHs,  without introducing any additional assumptions. The key feature is the  entropy, combined with basic principles of quantum physics. 

 The crucial point is to make a distinction between entropy, which measures the capacity of information storage, and the actual load of information (memory pattern) carried by a particular black hole. 
 The standard simplistic extrapolation of semiclassical result neglects the quantum backreaction as well as the effect of the information load. 
 In retrospect, this omission is striking since the information load, quantified by the MB parameter $\mu^{-1}$, can be enormous. The MB parameter $\mu^{-1}$, which is a new macroscopic characteristic of a BH, has the crucial advantage that it is invariant under the choice of basis used to describe the information content of the BH.

MB implies that the information stored in a BH can drastically influence its evolution~\cite{Dvali:2018xpy,Dvali:2018ytn,Dvali:2020wft, Dvali:2021tez, Dvali:2024hsb,Dvali:2025ktz}. Its \emph{swift} realization demonstrates that even the classical dynamics of a BH can be modified when the BH is perturbed, affecting, in particular, the frequency of gravitational waves (GWs) -- \cf
\eq (\ref{DeltaF1}) --
emitted during BH mergers~\cite{Dvali:2025sog}. This prediction has recently been explored in the context of GW observations to constrain MB~\cite{Yuan:2025hls}.

In this note, we investigate two questions: 
\begin{enumerate}
\item We derive observational signatures 
of swift MB in terms of $\mu^{-1}$ and $p$;
\item We study how $\mu^{-1}$ is determined by 
various collapsing sources. 
\end{enumerate}
Regarding the first item, we have elaborated on the predictions of swift MB for quasinormal modes (QNMs). Our main result is the frequency shift given in \eq \eqref{masterFormula}, which is a function of the critical exponent $p$ and the memory load $\mu^{-1}$.
With this we reproduced the earlier result of~\cite{Dvali:2025sog} but generalized it to arbitrary sign of the frequency shift.  
In particular, we have argued that $p$ determines whether the QNM frequency is increased or decreased.

Regarding the second item, one of the important points is that the memory load of a BH can largely exceed the 
actual information content carried by it.
 In particular, this is evident from \eq~\eqref{BHf}, which describes a state in which, while the  memory modes are highly entangled and information is shuffled, the memory load is maximal.  As a result, the BH formed out of (almost) featureless sources, such as the two colliding quanta, carry MB close to a maximum~(\ref{MaxMB}). 
 At the same time, the sources with  high diversity of features  necessarily lead to large memory loads.
 
Moreover, it is important that the MB
parameter $\mu^{-1}$ is not directly determined by the diversity of the features of the collapsing  source but also by the efficiency of the encoding mechanism translating these features into the pattern of the BH memory modes.  In particular, $\mu^{-1}$ can vary depending on how significantly the correlations among the source's features can reduce the usage of the BH memory space.  

As an important example, we derive 
bounds of $\mu^{-1}$ for BH formed by a collapsing star. While PBHs can carry a nearly maximal memory load \eqref{MaxMB}~\cite{Dvali:2025sog}, astrophysical BHs formed through stellar collapse are likely characterized by significantly smaller values of $\mu^{-1}$, reflecting the comparatively limited diversity of different stars.
This opens the intriguing possibility of using GW data to discriminate between BHs of primordial origin, early stellar populations, and more conventional astrophysical progenitors.

Finally, we emphasize that our analysis is perturbative, and thus valid only as long as the effect of MB remains a small correction. In the strong MB regime, the GW signal is likely to be substantially modified. This is especially relevant during the nonlinear merger phase of a BH coalescence, when the GW signal reaches its maximum amplitude, potentially altering, among other aspects, the inferred masses of merging BHs and even the template-based identification of GW events themselves.

In summary, swift MB suggests that the memory load of BHs is a new 
macroscopic property that influences the BH merger dynamics. Consequently, GW observations can provide a new probe of BH formation history and, more fundamentally, of the memory loaded into a BH. This makes the search for signatures of MB in BH mergers a particularly compelling target for GW astronomy.

\begin{acknowledgments}

It is a pleasure to thank Lasha Berezhiani and Otari Sakhelashvili for discussions. This work was supported in part by the Humboldt Foundation under Humboldt Professorship Award, by the European Research Council Gravities Horizon Grant AO number: 850 173-6,
by the Deutsche Forschungsgemeinschaft (DFG, German Research Foundation) under Germany's Excellence Strategy - EXC-2111 - 390814868, and Germany's Excellence Strategy under Excellence Cluster Origins. 

\paragraph*{Disclaimer} Funded by the European Union ERC, GRAVITES, 101071779. Views and opinions expressed are however those of the author(s) only and do not necessarily reflect those of the European Union or the European Research Council. Neither the European Union nor the granting authority can be held responsible for them.

\end{acknowledgments}

\bibliography{Refs,RefsManual}

\end{document}